\documentclass[%
 reprint,
 amsmath,amssymb,
 aps,
prl,
]{revtex4-1}

\usepackage{graphicx}
\usepackage{dcolumn}
\usepackage{bm}
\usepackage{comment}
\usepackage{braket}
\usepackage{color}
\usepackage{pstricks,pst-grad,color}

\begin{document}

\preprint{APS/123-QED}

\title{Equivalence of the effective non-hermitian Hamiltonians in the context of open quantum systems and strongly-correlated electron systems}
\author{Yoshihiro Michishita}
\email{michishita.yoshihiro.56e@st.kyoto-u.ac.jp}
 \affiliation{Department of Physics, Kyoto University, Kyoto 606-8502, Japan}
\author{Robert Peters}%
 \email{peters@scphys.kyoto-u.ac.jp}
 \affiliation{Department of Physics, Kyoto University, Kyoto 606-8502, Japan}

\date{\today}

\begin{abstract}
Recently, it has become clear that non-hermitian phenomena can be observed not only in open quantum systems experiencing gain and loss but also in equilibrium single-particle properties of strongly correlated systems.  However, the circumstances and requirements for the emergence of non-hermitian phenomena in each field are entirely different.  While the implementation of postselection is a significant obstacle to observe non-hermitian phenomena in open quantum systems, it is unnecessary in strongly correlated systems. 
   Until now, a relation between both descriptions of non-hermitian phenomena has not been revealed. In this paper, we close this gap and demonstrate that the non-hermitian Hamiltonians emerging in both fields are identical, and we clarify the conditions for the emergence of a non-hermitian Hamiltonian in strongly correlated materials. Using this knowledge, we propose a method to analyze non-hermitian properties without the necessity of postselection by studying specific response functions of open quantum systems and strongly correlated systems.
\end{abstract}

\maketitle

{\it{Introduction}}-- Recently, phenomena described by an effective non-hermitian Hamiltonian(NHH) are intensively studied, especially in the context of open quantum systems(OQS)\cite{PhysRevLett.120.146402,PhysRevB.56.8651, PhysRevLett.108.173901,brandstetter2014reversing, san2016majorana, doppler2016dynamically,PhysRevLett.116.133903,ashida2017parity,chen2017exceptional,feng2017non,PhysRevLett.120.146402, PhysRevX.8.031079,PhysRevLett.121.203001}.
Effective  NHH can induce novel topological phases\cite{PhysRevLett.120.146402,PhysRevX.8.031079,PhysRevLett.121.026808,PhysRevLett.121.086803,PhysRevLett.121.136802,kawabata2018symmetry,kawabata2019topological} and unique phenomena such as anomalous edge states\cite{PhysRevLett.116.133903,PhysRevLett.121.086803}, skin effects\cite{helbig2019observation,ghatak2019observation,xiao2019observation,hofmann2019reciprocal,okuma2019topological},  unusual quantum critical phenomena\cite{PhysRevLett.121.203001,ashida2017parity,yamamoto2019theory}, unidirectional invisibility \cite{PhysRevLett.106.213901,regensburger2012parity,feng2014single}, chiral transport \cite{PhysRevLett.86.787,peng2014parity,gao2015observation,doppler2016dynamically,xu2016topological}, and enhanced sensitivity \cite{PhysRevX.4.041001,PhysRevLett.112.203901,PhysRevLett.117.110802,hodaei2017enhanced,chen2017exceptional,yoon2018time,lau2018fundamental}. Although the total Hamiltonian is hermitian, the dynamics of the partial system alone can be described by an effective NHH, if the observed particle number of the partial system does not change during the time-evolution. An unchanged particle number in the partial system can be achieved by applying postselection. 
However, postselection becomes exceedingly difficult because the probability of finding a system with an unchanged particle number decreases exponentially\cite{SM}.
Thus, the study of non-hermitian phenomena in OQS has been limited to small systems.

Besides experimental and theoretical studies of effective NHH in the context of OQS, Kozii and Fu\cite{kozii2017non} demonstrated that an effective NHH describes the single-particle properties in strongly-correlated systems(SCS), which can result in the emergence of exceptional points and Fermi arcs in the spectral function.
The spectral function or other response functions can be calculated by the single-particle Green's function, $G^R(\omega,\bm{k})=(\omega-\mathcal{H}_0(\bm{k})-\Sigma(\omega,\bm{k}))^{-1}$ , where $H_0$ is the non-interacting part of the Hamiltonian and $\Sigma(\omega,\bm{k})$ is the self-energy.
The self-energy is represented by a non-hermitian matrix describing the correlations between particles, where the imaginary part of the self-energy describes the decay of a single-particle excitation.
The single-particle Green's function can thus be written as $G^R(\omega,\bm{k})=(\omega-\mathcal{H}_{eff}(\omega,\bm{k}))^{-1}$ , where $\mathcal{H}_{eff}(\omega,\bm{k})=\mathcal{H}_0(\bm{k})+\Sigma(\omega,\bm{k})$ is an effective NHH.  
It has been shown that non-hermitian properties of the effective Hamiltonian are related to correlation effects\cite{PhysRevB.98.035141,PhysRevB.99.121101,michishita2019relation,PhysRevB.100.100405,PhysRevB.100.115124,matsushita2019disorderinduced} and might be used to explain controversially discussed phenomena, such as quantum oscillations in  topological Kondo insulators\cite{PhysRevLett.121.026403} or the pseudogap phase in high-Tc cuprates\cite{kozii2017non}. It is interesting to note that in the context of Green's functions in SCS, postselection, which is usually difficult to realize, is not necessary to detect non-hermitian phenomena.

Until now,  studies of effective NHH in the context of OQS and SCS are proceeding nearly independently from each other.  It is unclear whether a relation between the effective NHH in both fields exists, and why postselection is not necessary in the context of SCS, while it is a  big obstacle in experimental studies of non-hermitian phenomena in OQS.

In this paper, we demonstrate that the NHH describing the Green's function  is equal to the NHH describing a single particle coupled to the rest of particles acting as a bath under postselection. For this purpose, we analyze the dynamics of a single particle in the Hubbard model using the quantum master equation (QME) in the context of OQS. 
The equivalence of the NHH in the single-particle spectral function and in the QME makes it possible to study non-hermitian phenomena in OQS by analyzing certain response functions without applied postselection.
Our analysis furthermore reveals why  postselection is not necessary to observe non-hermitian phenomena in the context of single-particle Green's functions.

\begin{figure}[t]
\includegraphics[width=0.8\linewidth]{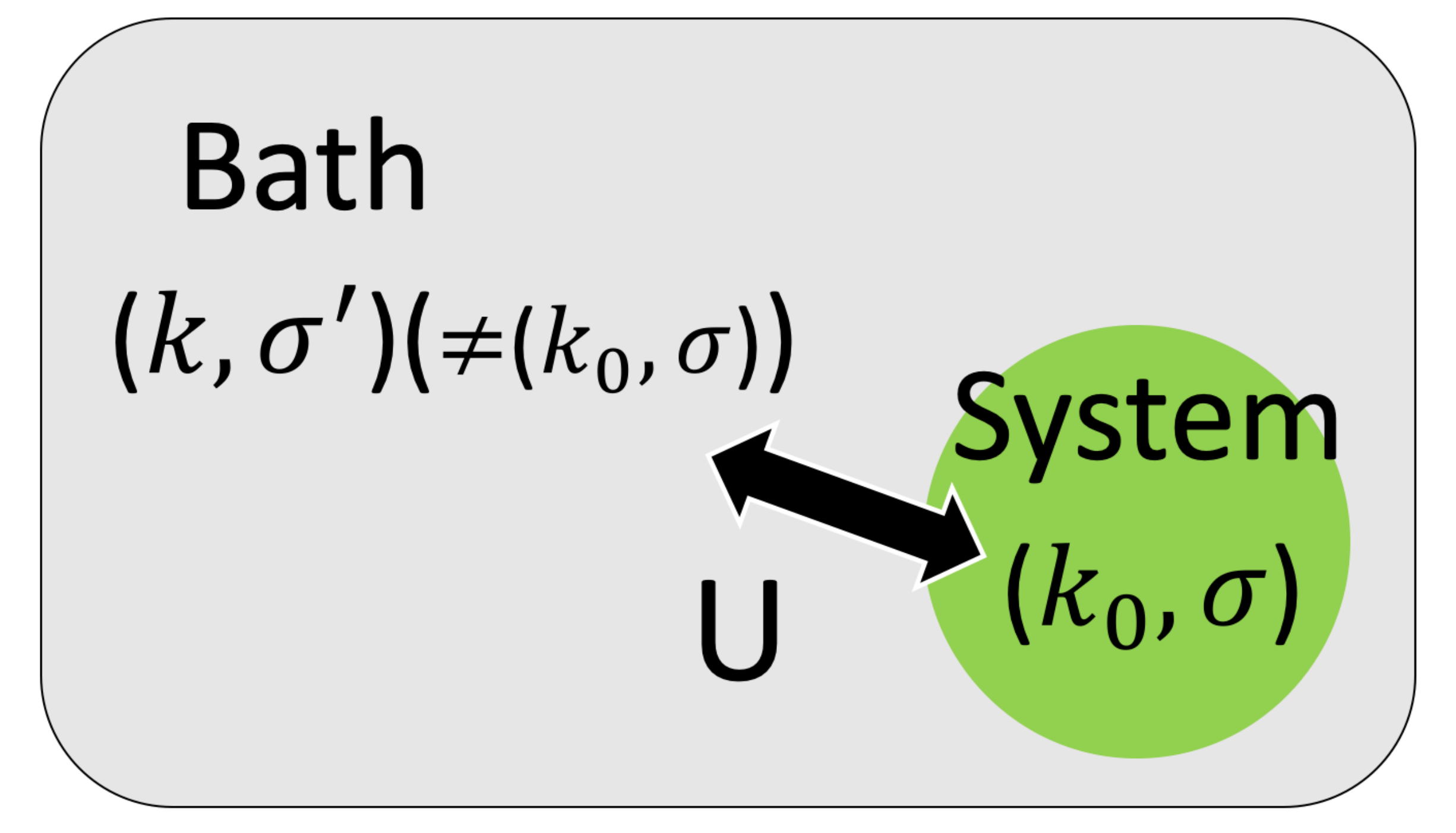}
\caption{To derive an effective non-hermitian Hamiltonian for the single-particle dynamics in the Hubbard model in the context of OQS, we divide the electrons into a system, including only one particle, and the rest of the particles, acting as bath.
\label{Fig1}}
\end{figure}

{\it{Quantum Master equation for the Hubbard model}} -- First, we derive the QME for the dynamics of a single particle in a strongly correlated material. Furthermore, we demonstrate that the effective NHH in the context of OQS under postselection corresponds to that in the single-particle Green's function in the context of SCS.
We here use the Hubbard model as a prototypical model describing SCS. 
In order to derive the effective NHH in the Hubbard model in the context of OQS, we divide the degrees of freedom into a system, describing a single particle, $(\bm{k_0},\sigma)$, at momentum $\bm{k_0}$ in spin-state $\sigma$, and a bath, which includes the rest of the electrons, see Fig. \ref{Fig1}. Thus, the total Hubbard Hamiltonian is divided into the Hamiltonian of the system, $\mathcal{H}_S$, the Hamiltonian of the bath, $\mathcal{H}_B$, and the coupling between system and bath, $\mathcal{H}_c$. The Hamiltonian becomes
\begin{eqnarray}
\mathcal{H}_{tot}&=&\sum_{\bm{k},\sigma}(\epsilon_{\bm{k}}+\mu_c)c^{\dagger}_{\bm{k}\sigma}c_{\bm{k}\sigma}\mathalpha{+}U \sum_i n_{i\uparrow}n_{i\downarrow}\label{X} \\
&=&\mathcal{H}_S+\mathcal{H}_B+\mathcal{H}_c\\
\mathcal{H}_S&=&(\epsilon_{\bm{k_0}}\mathalpha{+}\mu_c\mathalpha{+}Un_{\bar{\sigma}})c^{\dagger}_{\bm{k_0}\sigma}c_{\bm{k_0}\sigma}=\xi c^{\dagger}_{\bm{k_0}\sigma}c_{\bm{k_0}\sigma}\\
\mathcal{H}_B&=&\sum_{(\bm{k},\sigma^{\prime})\neq(\bm{k_0},\sigma)}(\epsilon_{\bm{k}}+\mu_c)c^{\dagger}_{\bm{k}\sigma^{\prime}}c_{\bm{k}\sigma^{\prime}}\\
&\mathalpha{+}&\frac{U}{N}\sum_{\sigma^\prime}\sum_{\substack{\bm{k_1},\bm{k_2},\\\bm{k_3},\bm{k_4}\\\neq(\bm{k_0},\sigma)}}
\!\!\!\!\!\!\delta_{\bm{k_1}+\bm{k_3},\bm{k_2}+\bm{k_4}}c^{\dagger}_{\bm{k_1}\sigma^{\prime}}
c_{\bm{k_2}\sigma^{\prime}}c^{\dagger}_{\bm{k_3}\bar{\sigma}^{\prime}}c_{\bm{k_4}\bar{\sigma}^{\prime}}\nonumber\\
\mathcal{H}_c&=&\frac{U}{N}\sum_{\substack{\bm{k_1},\bm{k_2},\bm{k_3}\\\neq\bm{k_0}}}\delta_{\bm{k_1}+\bm{k_3},\bm{k_0}+\bm{k_2}}\Bigl(c^{\dagger}_{\bm{k_0}\sigma}c_{\bm{k_1}\sigma}c^{\dagger}_{\bm{k_2}\bar{\sigma}}c_{\bm{k_3}\bar{\sigma}}+ h.c.\Bigr)\nonumber\\
&=&\frac{U}{N}\Bigl(\mathcal{C}_{\sigma}^{\dagger}\otimes\mathcal{B}_{\sigma}+h.c.\Bigr)\\
\mathcal{C}_{\sigma}&=&c_{\bm{k_0}\sigma}\\
\mathcal{B}_{\sigma}&=&\sum_{\substack{\bm{k_1},\bm{k_2},\bm{k_3}\\\neq\bm{k_0}}}\delta_{\bm{k_1}+\bm{k_3},\bm{k_0}+\bm{k_2}}c_{\bm{k_1}\sigma}c_{\bm{k_2}\bar{\sigma}}^{\dagger}c_{\bm{k_3}\bar{\sigma}},
\end{eqnarray}
where $c^{(\dagger)}_{\bm{k}\sigma}$ is an annihilation(creation) operator of an electron in momentum $\bm{k}$ and spin-direction $\sigma$. $\epsilon_{\bm{k}}$ is the energy dispersion, $\mu_c$ is the chemical potential, $U$ is the Hubbard interaction, and $N$ is the number of the lattice sites. Note that the coupling between the system and the bath corresponds to  a part of the two-particle interaction.

Starting from the von Neumann equation for the density matrix of the full system, $\frac{d}{dt}\rho(t)=-i[H,\rho(t)]$, we 
derive the QME for the density matrix of the system in second-order perturbation in  $\mathcal{H}_c$,
\begin{equation}
\frac{\partial}{\partial t}\rho^I_S(t)=-\int_{t_0}^{t}ds\mathrm{tr_B}\Bigl[ \mathcal{H}_c(t),\bigl[\mathcal{H}_c(s),\rho^I_S(s)\otimes \rho_B\bigr]\Bigr],\label{EQ:QME}
\end{equation}
where $\rho^I_S(t)$ is the density matrix of the system, i.e. the single particle. We here use the interaction representation
$\rho^I(t)=e^{i\mathcal{H}_St}\rho(t)e^{-i\mathcal{H}_St}$ and $\mathcal{H}_c(t)=e^{i(\mathcal{H}_S\otimes\mathcal{H}_B)t}\mathcal{H}_ce^{-i(\mathcal{H}_S\otimes\mathcal{H}_B)t}$.

The commutators in Eq.~(\ref{EQ:QME}) include terms such as\cite{SM} 
\begin{figure}[t]
\includegraphics[width=0.92\linewidth]{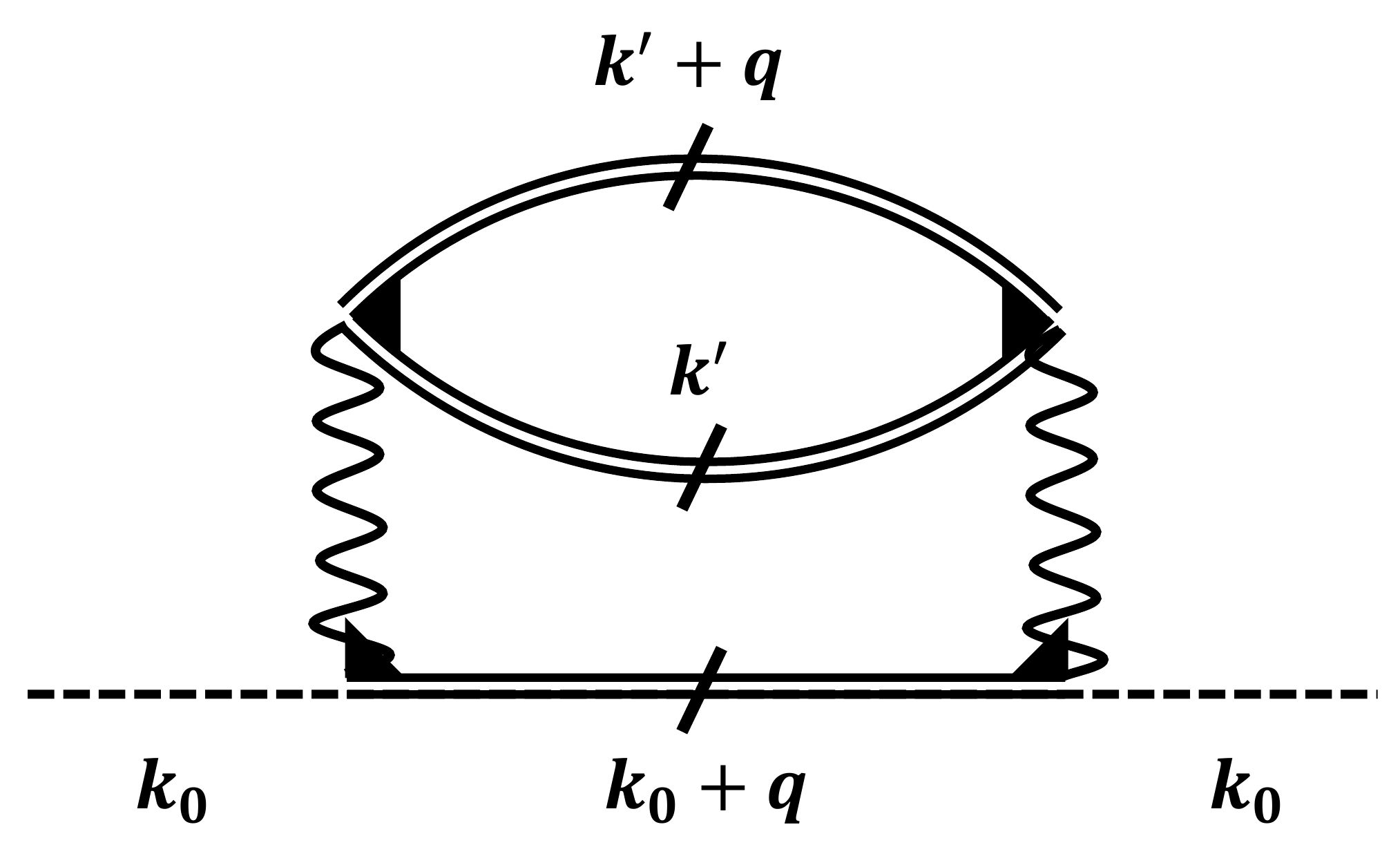}
\caption{Feynman diagram which describes the dynamics of the QME in second-order\label{fig:Sec_order_FD}. The slashed double lines correspond to full Green's function which  do not include the scattering to $\bm{k_0}$. The black triangle corresponds to the full two-particle vertex, which does not include scattering via $\bm{k_0}$.}
\end{figure}
\begin{eqnarray}
&&\mathcal{C}^{\dagger}_{\sigma}\mathcal{C}_{\sigma}\rho_S(s)\otimes\mathrm{Tr_B}\Bigl[\mathcal{B}_{\sigma}(t)\mathcal{B}^{\dagger}_{\sigma}(s)\rho_B\Bigr]=\nonumber\\
 && \mathcal{C}^{\dagger}_{\sigma}\mathcal{C}_{\sigma}\rho_S(s)\otimes\mathrm{Tr_B}\Bigl[\sum_{\bm{k_1},\bm{k_2},\bm{k_3}}\delta_{\bm{k_1}+\bm{k_3},\bm{k_0}+\bm{k_2}}\\
 &&\ \ \ \times c_{\bm{k_1}\sigma}(t)c_{\bm{k_2}\bar{\sigma}}^{\dagger}(t)c_{\bm{k_3}\bar{\sigma}}(t)c^{\dagger}_{\bm{k_3}\bar{\sigma}}(s)c_{\bm{k_2}\bar{\sigma}}(s)c_{\bm{k_1}\sigma}^{\dagger}(s)\rho_B\Bigr].\nonumber
\end{eqnarray}
This trace over three creation and three annihilation operators including the time evolution by
 the full Hamiltonian, only missing the scattering via ($\bm{k_0},\sigma$), appears in the second-order diagram for the self-energy shown in Fig.~\ref{fig:Sec_order_FD}.

Because the amplitude of a single scattering process via $\bm{k_0}$ vanishes in the limit of an infinite large bath, $N\rightarrow \infty$,  the self-energy shown in Fig.~\ref{fig:Sec_order_FD} becomes the exact self-energy in second-order perturbation in $\mathcal{H}_c$(not $U$). 
Even when considering higher-order perturbations in $\mathcal{H}_c$, we find that the QME still can be described by the self-energy\cite{SM}.
Collecting all terms in Eq.(\ref{EQ:QME}), we obtain
\begin{eqnarray}
\frac{\partial}{\partial t}\rho^I_S(t)&=&\int_{t_0}^t ds\biggl[
-i \mathrm{Re}(S_{l}(t-s))[\mathcal{C}^{\dagger}_{\sigma}\mathcal{C}_{\sigma},\rho^I_S(s)]\label{QMC_exact}\\
&&+ i\mathrm{Re}(S_{g}(t-s))[\mathcal{C}_{\sigma}\mathcal{C}^{\dagger}_{\sigma},\rho^I_S(s)]\nonumber\\
&& + \mathrm{Im}(S_{l}(t-s))\Bigl(\{\mathcal{C}^{\dagger}_{\sigma}\mathcal{C}_{\sigma},\rho^I_S(s)\}-2\mathcal{C}_{\sigma}\rho^I_S(s)\mathcal{C}^{\dagger}_{\sigma}\Bigr)\nonumber\\
&& +\mathrm{Im}( S_{g}(t-s))\Bigl(\{\mathcal{C}_{\sigma}\mathcal{C}^{\dagger}_{\sigma},\rho^I_S(s)\}-2\mathcal{C}^{\dagger}_{\sigma}\rho^I_S(s)\mathcal{C}_{\sigma}\Bigr)\biggr]\nonumber,
\end{eqnarray}
with
\begin{eqnarray}
S_l(t) &=& \Sigma^T(t)e^{i\xi t}\nonumber\\
S_g(t) &=&( \Sigma^R(t)-\Sigma^T(t)\bigr)e^{i\xi t}\nonumber
\end{eqnarray}
where $\Sigma^T$ is the time-ordered self-energy, $\Sigma^R$ is the retarded self-energy, and $\xi=\epsilon_{\bm{k_0}}\mathalpha{+}\mu_c\mathalpha{+}Un_{\bar{\sigma}}$.

We see that the time-evolution of the density-matrix of a single particle at $(\bm{k_0},\sigma$) is governed by the self-energy $\Sigma^{R/T}_{k_0,k_0}(s)$. However, because Eq.~(\ref{QMC_exact}) includes gain and loss terms, i.e. $2\mathcal{C}_{\sigma}\rho^I_S(s)\mathcal{C}^{\dagger}_{\sigma}$ and $2\mathcal{C}^{\dagger}_{\sigma}\rho^I_S(s)\mathcal{C}_{\sigma}$, the dynamics cannot be described by an effective NHH alone.

We next fix the particle number of the system, which corresponds to applying postselection. We restrict the Hilbert space to states where $c_{\bm{k_0}\sigma}^{\dagger}c_{\bm{k_0}\sigma}+c_{\bm{k_0}\bar{\sigma}}^{\dagger}c_{\bm{k_0}\bar{\sigma}}=1$. We furthermore assume the absence of magnetism, which results in
 $\ c_{\bm{k_0}\sigma}^{\dagger}c_{\bm{k_0}\sigma}=c_{\bm{k_0}\sigma}c_{\bm{k_0}\sigma}^{\dagger}$ in the restricted Hilbert space. Due to these restrictions, the gain and loss terms vanish in Eq.~(\ref{QMC_exact}), and the commutators and anticommutators can be summed up
\begin{eqnarray}
&&\frac{\partial}{\partial t}\rho^{I \ PS}_S(t)=\label{PS_QME}\\
&&-i\int_{t_0}^t ds \Bigl( \mathcal{S}_{eff}(t-s)\rho^{I \ PS}_S(s) - \rho^{I \ PS}_S(s) \mathcal{S}_{eff}^{\dagger}(t-s)\Bigr)\nonumber
\end{eqnarray}
\begin{eqnarray}
\mathcal{S}_{eff}(t-s)&=&\Sigma^R(t-s)e^{i\xi(t-s)}c_{\bm{k_0}\sigma}^{\dagger}c_{\bm{k_0}\sigma},
 \end{eqnarray}
where $\rho^{(I) \ PS}_S(t)$ is the density matrix with applied postselection.
By using the Markov approximation, $\rho_S(s)\rightarrow\rho_S(t)$, and taking the limit $t_0\rightarrow-\infty$, we find that the density matrix of a single particle under postselection can be written as
\begin{eqnarray}
\frac{\partial}{\partial t}\rho^{PS}_S(t) &=& -i \Bigl( \mathcal{H}_{eff}\rho^{PS}_S(t) - \rho^{PS}_S(t) \mathcal{H}_{eff}^{\dagger}\Bigr)\label{MD1}\\
\mathcal{H}_{eff}&=& \mathcal{H}_0+\Sigma^R_{0}(\xi)c_{\bm{k_0}\sigma}^{\dagger}c_{\bm{k_0}\sigma},\label{MD2}
\end{eqnarray}
which corresponds to the von Neumann equation with an effective NHH.
Thus,  the time-evolution of the density-matrix of a single particle $(\bm{k_0},\sigma)$ is given by an effective NHH including the self-energy, if postselection is applied.
 We note that the frequency dependence of the self-energy has vanished because of the Markov approximation and taking the limit $t_0\rightarrow-\infty$.

However, in the context of SCS, the Green's function is described by an effective NHH without postselection\cite{kozii2017non,PhysRevB.98.035141,PhysRevB.99.121101,michishita2019relation,PhysRevB.100.100405,PhysRevB.100.115124,matsushita2019disorderinduced}.
To clarify the reason why postselection is not necessary in this context, we will now introduce the retarded Green's function in the steady state using the density matrix form, which is given as $G^R_{OQS}(t)=-i\Theta(t)\mathrm{Tr}\Bigl[\bigl(\mathcal{C}(t)\mathcal{C}^{\dagger}(0)+\mathcal{C}^{\dagger}(0)\mathcal{C}(t)\bigr)\rho^{SS}_S\Bigr]$.
Here, $\rho^{SS}_S$ is the density matrix of the single particle in the long-time limit (steady-state).
Combining the density-matrix, $\rho^{SS}_S$, with the creation-operator, $\mathcal{C}^{\dagger}$, we define the density-matrix describing the single-particle Green's function, $\rho^{RGF}_S=\mathcal{C}^{\dagger}\rho^{SS}_S+\rho^{SS}_S\mathcal{C}^{\dagger}$. 
Thus, we can rewrite the Green's function as 
\begin{displaymath}
G^R_{OQS}(t) = -i\Theta(t)\mathrm{Tr}\bigl[\mathcal{C}\rho^{RGF}_S(t)\bigr],\end{displaymath}
 where the time evolution of $\rho^{RGF}_S(t)$ is given by the QME in Eq.~(\ref{QMC_exact}).

When considering a system which includes only a single particle, ($\bm{k_0},\sigma$), $\rho^{RGF}_S(t)$ is given by the following matrix element, $\ket{\sigma}\bra{0}$, where $\ket{\sigma}=c^{\dagger}_{\bm{k_0},\sigma}\ket{0}$. Gain and loss terms vanish in the time evolution for this matrix element, because $\mathcal{C}^{\dagger}\ket{\sigma}\bra{0}\mathcal{C}=\mathcal{C}\ket{\sigma}\bra{0}\mathcal{C}^{\dagger}=0$.
Therefore,  the QME can be written as  
\begin{eqnarray}
 \frac{\partial}{\partial t}\rho^{I \ RGF}_S(t)
&=& -i\int_{t_0}^t ds \Bigl( \mathcal{S}_{eff}(t-s)\rho^{I \ RGF}_S(s) \label{GF_QME}\\
&&- \rho^{I \ RGF}_S(s) \mathcal{S}_{eff}^{\dagger}(t-s)\Bigr)\nonumber\\
\Rightarrow -i\omega \rho^{RGF}_S(&\omega&) - \rho^{RGF}_{S}(t_0)=\\ && -i\Bigl( \mathcal{H}_{eff}(\omega)\rho^{RGF}_S(\omega) - \rho^{RGF}_S(\omega) \mathcal{H}_{eff}^{\dagger}(\omega)\Bigr)\nonumber\\
\mathcal{H}_{eff}(\omega) &=& \mathcal{H}_0+\Sigma^R(\omega)c^{\dagger}_{\bm{k_0},\sigma}c_{\bm{k_0},\sigma}
\end{eqnarray}
and the Green's function becomes
\begin{eqnarray}
G^R_{OQS}(\omega)& =& -i\mathrm{Tr}\bigl[\mathcal{C}\rho^{RGF}_S(\omega)\bigr] = \frac{1}{\omega-\xi-\Sigma^R(\omega)}\label{NMD}
\end{eqnarray}

We here have demonstrated the following statements:
First, the Green's function of a single particle described as an OQS and its effective NHH is identical to the Green's function and its NHH in closed equilibrium systems.
Second, the dynamics of $\rho^{PS}_S$ and $\rho^{RGF}_S$ are described by the same equations, Eq.~(\ref{PS_QME}) and Eq.~(\ref{GF_QME}). 
We can conclude that the effective NHH describing the dynamics under postselection is identical to the effective NHH describing the Green's function in SCS.
Thus, we can analyze non-hermitian phenomena, which are observable in OQSs under  postselection, by studying the spectral function $A(\omega)=-\frac{1}{\pi}\mathrm{Im}G^R_{OQS}(\omega)$ in equilibrium or the nonequilibrium steady state. While postselection becomes increasingly difficult in large systems, the analysis of spectral properties remains feasible. We note that non-hermitian properties may occur in different response functions than the single-particle spectral function and that the correspondence between the NHH in the density matrix under postselection and the NHH in the response function depends on the kind of the postselection.
Third, because the density matrix describing the Green's functions in the context of OQS is given  by the off-diagonal matrix element, i.e. $\ket{\uparrow}\bra{0}$, gain and loss terms vanish in the QME, and postselection is unnecessary to derive an effective NHH.
We note that, even if we consider larger systems, for example a system including ($\bm{k_0},\uparrow$) and ($\bm{k_0},\downarrow$), gain and loss contributions in the QME for the Green's function  vanish \cite{SM}.

{\it{Dynamics of the Hubbard model in the quantum Master equation}}
 -- 
 Finally, we use the above-introduced QME to describe single-particle properties in the Hubbard model on a 2D square lattice.
 We furthermore show that the Markov approximation, which ignores the memory effect of the QME dynamics, fails to describe the full spectral function in the Mott phase of the Hubbard model in which non-Markovian dynamics plays an important role.
 
We have shown above that the time-evolution of the density matrix is determined by the self-energy in the QME. We here use the dynamical mean field theory (DMFT) combined with the numerical renormalization group (NRG) to calculate an approximate self-energy.\cite{RevModPhys.68.13,RevModPhys.80.395,PhysRevB.74.245114}
Using the self-energy obtained by DMFT/NRG in the QME, Eq.~(\ref{QMC_exact}),  we show the relaxation dynamics of the density matrix into the steady state, and demonstrate that the spectral function calculated by the QME approach is identical with the spectral function directly obtained by DMFT/NRG.

\begin{figure}[t]
\includegraphics[width=0.98\linewidth]{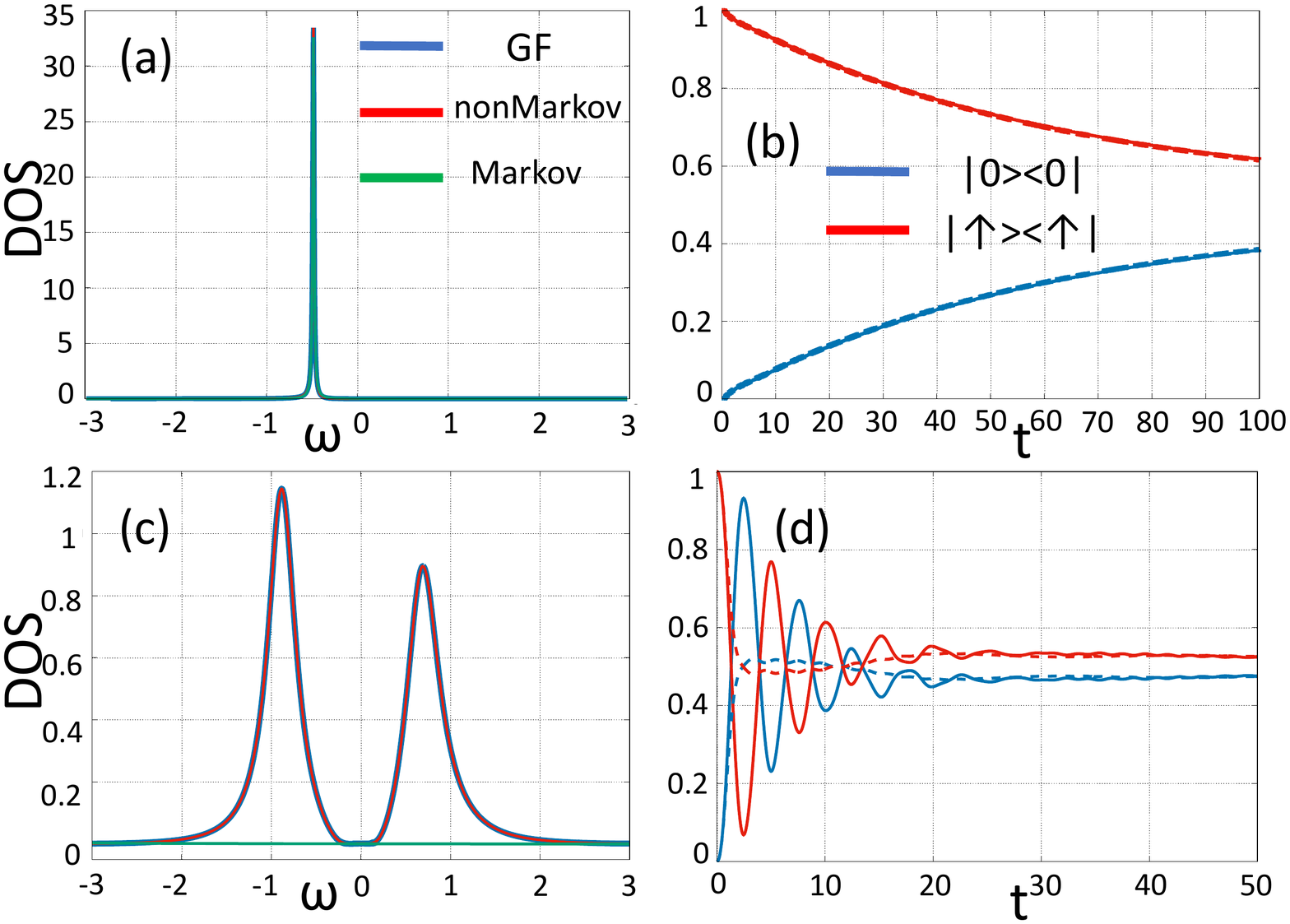}
\caption{Spectral function and the time-evolution of the diagonal elements into the steady state in the weak-coupling regime and in the Mott insulating phase.\label{fig:MottSF1}
The parameters in Fig.~(a) and (b) are as follows: $\epsilon_k=-0.49$
$\mu_c=-0.2$, $U=0.4$, and the temperature $T=0.001$. The parameters in Fig.~(c) and (d) are as follows;  $\epsilon_k=-0.12$, $t=0.1$, $\mu_c=-0.8$, $U=1.6$,$T=0.00006$. The blue, red and green lines in (a) and (c) show the spectral function as calculated by the Green's function, non-Markov QME (Eq.~\ref{GF_QME}), and the Markov QME in the limit of $t_0 \rightarrow -\infty$ (Eq.~(\ref{MD1})), respectively. 
The blue and the red lines in (b) and (d) show the dynamics of the diagonal elements $|0><0|$ and $|\uparrow><\uparrow|$ from the initial state $\rho_{i}=|\uparrow><\uparrow|$. The full lines and the dashed lines correspond to the non-Markovian dynamics and the Markovian dynamics, respectively.}
\end{figure}

In Fig. \ref{fig:MottSF1}, we compare the spectral functions calculated by the QME and by DMFT/NRG for the weak-coupling regime (Fig. \ref{fig:MottSF1}(a)) and the Mott insulator (Fig. \ref{fig:MottSF1}(c)) for  $\bm{k_0}=(0.4\pi,0.4\pi)$. We furthermore include a comparison between the QME approach using the Markov approximation and the full dynamics. In the weak-coupling regime, the spectral functions obtained by DMFT and the QME with and without Markov approximation agree with each other. 
Figure \ref{fig:MottSF1}(b) shows the time-evolution of the diagonal elements of the density matrix with and without Markov approximation in the QME, Eq. (\ref{QMC_exact}). In the weak-coupling regime, memory effects are not important and therefore the Markov approximation works well. The dynamics without memory effects is given by an exponential decay as shown in Fig.~\ref{fig:MottSF1}(b).
We conclude that that the Markov approximation can describe the full dynamics of the system in the weak-coupling regime, Fig.~\ref{fig:MottSF1}(a-b).

In the Mott-insulating phase, shown in Fig.~\ref{fig:MottSF1}(c), the non-Markov spectral function does also agree with the spectral function obtained by DMFT/NRG.  On the other hand, the spectral function calculated with the Markov approximation is nearly zero.
In the Mott insulating regime, the Markov approximation describes strong dissipation due to the strong scattering with the bath electrons and the resulting spectral function has only a small and  wide peak.
Non-Markovian dynamics is essential to correctly describe the strongly interacting system. Both peaks in the spectral function are described by quasi-particles which follow non-Markovian dynamics. In Fig.~\ref{fig:MottSF1}(d), we show the dynamics of the diagonal elements of the density matrix comparing between Markovian and non-Markovian dynamics. Both approaches show a strong decay into the same steady state. Additional to the strong decay of the matrix element of the density-matrix, the non-Markovian dynamics show a strong oscillatory behavior of the occupation number. 

In the supplemental materials\cite{SM}, we perform a similar analysis for the periodic Anderson model, showing that also in this model the spectral function of a small system described as an OQS and its effective NHH are identical to the Green's function and its NHH in closed equilibrium systems.

{\it{Summary and Discussion}}
-- By analyzing the Hubbard model as an OQS, we have proved that the effective NHH appearing in the context of OQS and equilibrium Green's functions are identical.  
We have demonstrated that the spectral function of a single particle described as an OQS is given by the same non-hermitian Hamiltonian describing the density matrix of the particle under postselection. Thus, non-hermitian phenomena that have been analyzed in the dynamics of a system under postselection can also be studied by analyzing spectral functions both in OQS and SCS without postselection. We have also shown that postselection is not necessary to derive a NHH from the spectral function, because off-diagonal elements govern the dynamics of the spectral function, and gain and loss contributions automatically vanish.
In the process of deriving the QME for a single particle, we have succeeded in showing that Feynman diagrams, e.g.,  representing the self-energy, describe the non-Markovian dynamics of a fermionic system coupled to a fermionic bath. This technique might also be applied to other systems, such as QuBits coupled to fermionic baths.
Finally, we have demonstrated the importance of non-Markovian dynamics to describe the dynamics in the strongly correlated regime.

{\it{Acknowledgments}}
 -- YM thanks C. Uchiyama, I. Hashimoto, K.Mizuta, K.Takasan, and N.Hatano for fruitful discussion and valuable comments. This work is partly supported by JSPS KAKENHI Grants No. JP18H04316 and No. JP18K0351. Computer simulations were performed on the supercomputer of the ISSP in the University of Tokyo.

\bibliography{paper}
\clearpage

\renewcommand{\thesection}{S\arabic{section}}
\renewcommand{\theequation}{S\arabic{equation}}
\renewcommand{\thefigure}{S\arabic{figure}}

\setcounter{equation}{0}
\setcounter{figure}{0}

\onecolumngrid
\begin{center}
{\large
{\bfseries Supplemental Materials for \\ ``Equivalence of the effective non-hermitian Hamiltonians in the context of open quantum systems and strongly-correlated electron systems'' }}
\end{center}

\vspace{10pt}

\twocolumngrid

\section{S1. \ \ \label{app:PS}Postselection in open quantum systems}

Repeated measurement of the bath, selecting a specific class of outcomes, and resetting the bath to its initial state, has a strong effect on the dynamics of the density matrix of the system.  
 This is known as postselection and can be written as
\begin{equation}
\rho_S \rightarrow \rho^{\prime}_S = \mathcal{P}_m\rho_S\mathcal{P}_m/\mathrm{tr}[\mathcal{P}_m\rho_S\mathcal{P}_m]
\end{equation}
where $\rho_S$ is the density matrix of the system and $\mathcal{P}_m$ is the projection operator on the result of the measurement, $m$. 
The dynamics of the projected density matrix of the system under postselection becomes
\begin{eqnarray}
 \rho^{\prime}_S(t+\delta t) & =& \rho^{\prime}_S(t) -i \Bigl( \mathrm{H}^{\prime}_{eff} \rho^{\prime}_S(t)- \rho^{\prime}_S(t) \mathrm{H}^{\prime\dagger}_{eff}\\&& - \sum_{\alpha}\gamma_{\alpha}\mathcal{L}^{\prime}_{\alpha} \rho^{\prime}_S(t) \mathcal{L}^{\prime\dagger}_{\alpha}\Bigr)\\
\rho^{\prime}_S(t+\delta t) &=& \mathcal{P}_n\rho_S(t+\delta t)\mathcal{P}_n/\mathrm{tr}[\mathcal{P}_n\rho_S\mathcal{P}_n]\\
\mathcal{O}^{\prime} &=& \mathcal{P}_n \mathcal{O} \mathcal{P}_m.
\end{eqnarray}
where $\mathcal{O}$ is an arbitrary operator acting on the system and $\gamma_{\alpha}\mathcal{L}^{\prime}_{\alpha} \rho^{\prime}_S(t) \mathcal{L}^{\prime\dagger}_{\alpha}$ describes the gain or loss, resulting in a change of the particle number of the system.
When the repeated measurement of the system is performed in a way so that the particle number of the system does not change, the gain and loss modes disappear due to the projection operators. Thus, the dynamics of the open quantum system under postselection is described by an effective non-hermitian Hamiltonian.

\section{S2. \ \ \label{app:NZE}Representation of the dynamics in the quantum master equation by the self-energy}
\subsection{A. \ \ \label{app:TCL}Exact representation for open quantum systems by projection operator method}
Here, we review the exact master equation of the dynamics of open quantum systems by using the projection operator $\mathcal{P}$ and $\mathcal{Q}$. $\mathcal{P}$ is the projection operator on the Hilbert space, where the system and the bath are disentangled,  $\mathcal{P}\rho=\mathrm{tr_B}[\rho]\otimes \rho_B$ and $\mathcal{Q}=1-\mathcal{P}$.

We suppose that odd moments of the interaction, $\mathcal{H}_c$, which describes the coupling between bath and system, vanish. Thus, 
\begin{eqnarray}
\mathrm{tr_B}\Bigl[ \underbrace{\mathcal{H}_c(t_1)\dots\mathcal{H}_c(t_{2n+1})}_{\text{odd power}}\rho_B \Bigr]=0\label{appa},
\end{eqnarray}
which leads to the relation 
\begin{eqnarray}
 \mathcal{P} \mathcal{L}(t_1)\dots\mathcal{L}(t_{2n+1})\mathcal{P}&=&0, \label{assumption1}
 \end{eqnarray}
 where
 \begin{eqnarray}
 \mathcal{L}(t_n)\rho=-i[\mathcal{H}_c(t_n),\rho].
\end{eqnarray}

The dynamics of the disentangled system and its complement can be written as
\begin{eqnarray}
 \frac{\partial}{\partial t}\mathcal{P}\rho(t) &=& \mathcal{P}\frac{\partial}{\partial t}\rho(t)
 = \mathcal{P}\mathcal{L}(t) \Bigl(\mathcal{P}+ \mathcal{Q}\Bigr) \rho(t)\label{Pro5}\\
 \frac{\partial}{\partial t}\mathcal{Q}\rho(t) &=& \mathcal{Q}\frac{\partial}{\partial t}\rho(t)
 = \mathcal{Q}\mathcal{L}(t) \Bigl(\mathcal{P}+ \mathcal{Q}\Bigr) \rho(t).\label{Pro1}
\end{eqnarray}
Using Eq.~(\ref{Pro5}), Eq.~(\ref{Pro1}), and Eq.~(\ref{assumption1}), we can derive 
\begin{eqnarray}
 \mathcal{Q}\rho(t) &=& \mathcal{Q}\rho(t_0) + \int_{t_0}^{t}ds \mathcal{Q}\mathcal{L}(s) \Bigl(\mathcal{P}+ \mathcal{Q}\Bigr) \rho(s)\\
 &=& \mathcal{Q}\rho(t_0)+ \int_{t_0}^{t}ds \mathcal{Q}\mathcal{L}(s)\mathcal{Q}\rho(t_0)\label{Pro2}\\
 && \ + \int_{t_0}^{t}ds \mathcal{Q}\mathcal{L}(s)\mathcal{P}\rho(s)\nonumber\\
 && \ + \int_{t_0}^{t}dt_1 \mathcal{Q}\mathcal{L}(t_1)\int_{t_0}^{t_1}dt_2\mathcal{Q}\mathcal{L}(t_2)\Bigl(\mathcal{P}+ \mathcal{Q}\Bigr) \rho(t_2)\nonumber\\
 &=& \mathcal{G}(t,t_0)\mathcal{Q}\rho(t_0) + \int_{t_0}^{t}ds \mathcal{G}(t,s) \mathcal{Q}\mathcal{L}(s)\mathcal{P}\rho(s),\nonumber
\end{eqnarray}
where we introduce the forward propagator
\begin{eqnarray}
\mathcal{G}(t,s)= \mathrm{T} \exp{\Bigl[ \int_s^t ds^{\prime} \mathcal{Q}\mathcal{L}(s^{\prime})\Bigr]}.
\end{eqnarray}
$\mathrm{T}$ describes the chronological time ordering. By inserting Eq.(\ref{Pro2}) into Eq.(\ref{Pro5}), we can derive the dynamics of the system, which reads
\begin{eqnarray}
\frac{\partial}{\partial t}\mathcal{P}\rho(t) &=& \mathcal{I}(t,t_0)\mathcal{Q}\rho(t_0)+ \int_{t_0}^tds\mathcal{K}(t,s)\mathcal{P}\rho(s)\nonumber\\
\label{QME}\\
\mathcal{I}(t,t_0)&=&\mathcal{P}\mathcal{L}(t)\mathcal{G}(t,t_0)\mathcal{Q}\label{Pro7}\\
\mathcal{K}(t,s)&=&\mathcal{P}\mathcal{L}(t)\mathcal{G}(t,s)\mathcal{Q}\mathcal{L}(s)\mathcal{P}.\label{Pro6}
\end{eqnarray}
Eq.(\ref{QME}) is the exact quantum Master equation by using the projection operator, which is known as Nakajima-Zwanzig equation.\cite{breuer2002theory}

The first term in Eq.~(\ref{QME}) disappears when we assume that the system and the bath are not entangled in the initial state.
 Furthermore, if we change $\mathcal{P}\rho(s) \rightarrow \mathcal{P}\rho(t)$ in Eq.~(\ref{QME}), the dynamics of the system is determined only by the current state of the system, which is known as Markov approximation.

\subsection{B. \ \ \label{PT}Correspondence between the perturbation approach in Nakajima-Zwanzig equation and the diagram approach in Green function method }
In this section, we confirm that the dynamics in Eq.~(\ref{QME}) is described by the self-energy.

First, we confirm this result for the second order perturbation about the interaction $\mathcal{H}_c$,  which corresponds to a second order process in $\mathcal{L}$. We note that any odd order perturbation term disappears because of Eq.~(\ref{assumption1}). The second order term of $\mathcal{K}(t)$ reads,
\begin{eqnarray}
\mathcal{K}_2(t,s)&=& \mathcal{P}\mathcal{L}(t)\mathcal{L}(s)\mathcal{P}.\label{2order}
\end{eqnarray}
Applying this to the Hubbard model in the main text, we obtain
\begin{eqnarray}
\mathcal{K}_2(t,s)\rho_S(s)&=&- \mathrm{tr_B}\Bigl[ \mathcal{H}_c(t),\bigl[\mathcal{H}_c(s),\rho^I_S(s)\otimes \rho_B\bigr]\Bigr],\nonumber\\
\label{2order}
\end{eqnarray}
which becomes
\begin{eqnarray}
&&-\int_{0}^t ds \mathrm{tr_B}\Bigl[\mathcal{C}^{\dagger}_{\sigma}(t)\otimes\mathcal{B}_{\sigma}(t),\bigl[\mathcal{C}_{\sigma}(s)\otimes\mathcal{B}^{\dagger}_{\sigma}(s),\rho^I_S(s)\otimes\rho_B\bigr]\Bigr]\nonumber\\
&&+\text{h.c.}\nonumber
\end{eqnarray}
We here have used that the bath is in equilibrium so that $[\mathcal{H}_B,\rho_B]=0$ is satisfied.
Performing the commutators in the above expression, we find the following terms
\begin{widetext}
\begin{eqnarray}
\Bigl(\mathcal{C}^{\dagger}_{\sigma}(t)\mathcal{C}_{\sigma}(s)\rho^I_S(s)-\mathcal{C}_{\sigma}(s)\rho^I_S(s)\mathcal{C}^{\dagger}_{\sigma}(t)\Bigr)&\otimes&\mathrm{tr_B}\Bigl[\mathcal{B}_{\sigma}(t)\mathcal{B}^{\dagger}_{\sigma}(s)\rho_B\Bigr]\nonumber\\&=&\Bigl(\mathcal{C}^{\dagger}_{\sigma}\mathcal{C}_{\sigma}\rho^I_S(s)-\mathcal{C}_{\sigma}\rho^I_S(s)\mathcal{C}^{\dagger}_{\sigma}\Bigr)\otimes \Bigl( i\Sigma_2^T(t-s)e^{i\xi (t-s)}\Bigr)\label{Torder1}\\
\Bigl(\rho^I_S(s)\mathcal{C}^{\dagger}_{\sigma}(s)\mathcal{C}_{\sigma}(t)-\mathcal{C}_{\sigma}(t)\rho^I_S(s)\mathcal{C}^{\dagger}_{\sigma}(s)\Bigr)&\otimes&\mathrm{tr_B}\Bigl[\rho_B\mathcal{B}_{\sigma}(s)\mathcal{B}^{\dagger}_{\sigma}(t)\Bigr]\nonumber\\
&=& \Bigl(\rho^I_S(s)\mathcal{C}^{\dagger}_{\sigma}\mathcal{C}_{\sigma}-\mathcal{C}_{\sigma}\rho^I_S(s)\mathcal{C}^{\dagger}_{\sigma}\Bigr)\otimes \Bigl(i\Sigma_2^T(t-s)e^{i\xi (t-s)}\Bigr)^{\dagger}\label{Torder2}\\
\Bigl(\mathcal{C}_{\sigma}(t)\mathcal{C}^{\dagger}_{\sigma}(s)\rho^I_S(s)-\mathcal{C}^{\dagger}_{\sigma}(s)\rho^I_S(s)\mathcal{C}_{\sigma}(t)\Bigr)&\otimes&\mathrm{tr_B}\Bigl[\mathcal{B}^{\dagger}_{\sigma}(t)\mathcal{B}_{\sigma}(s)\rho_B\Bigr]\nonumber\\
&=&\Bigl(\mathcal{C}_{\sigma}\mathcal{C}^{\dagger}_{\sigma}\rho^I_S(s)-\mathcal{C}^{\dagger}_{\sigma}\rho^I_S(s)\mathcal{C}_{\sigma}\Bigr)\otimes\Bigl(i(\Sigma_2^R(t-s)-\Sigma_2^T(t-s))e^{i\xi (t-s)}\Bigr)^{\dagger}\label{Torder3}\\
\Bigl(\rho^I_S(s)\mathcal{C}_{\sigma}(s)\mathcal{C}^{\dagger}_{\sigma}(t)-\mathcal{C}^{\dagger}_{\sigma}(t)\rho^I_S(s)\mathcal{C}_{\sigma}(s)\Bigr)&\otimes&\mathrm{tr_B}\Bigl[\rho_B\mathcal{B}^{\dagger}_{\sigma}(s)\mathcal{B}_{\sigma}(t)\Bigr]\nonumber\\
&=&\Bigl(\rho^I_S(s)\mathcal{C}_{\sigma}\mathcal{C}^{\dagger}_{\sigma}-\mathcal{C}^{\dagger}_{\sigma}\rho^I_S(s)\mathcal{C}_{\sigma}\Bigr)\otimes \Bigl(i(\Sigma_2^R(t-s)-\Sigma_2^T(t-s))e^{i\xi (t-s)}\Bigr)\label{Torder4}
\end{eqnarray}
\end{widetext}

We note that terms such as $\mathrm{tr_B}[\mathcal{B}_{\sigma}(t)\mathcal{B}^{\dagger}_{\sigma}(s)\rho_B]$ can be visualized as shown in Fig. \ref{fig:Sec_order_FD} and correspond to the $\Sigma_2^T(t-s)$ (time-ordered self-energy) in second-order. This correspondence is shown in more detail in S3.
\begin{figure}[h]
\includegraphics[width=0.92\linewidth]{Sec_Order_FD_new.pdf}
\caption{Second order Feynman diagram\label{fig:Sec_order_FD}}
\includegraphics[width=0.92\linewidth]{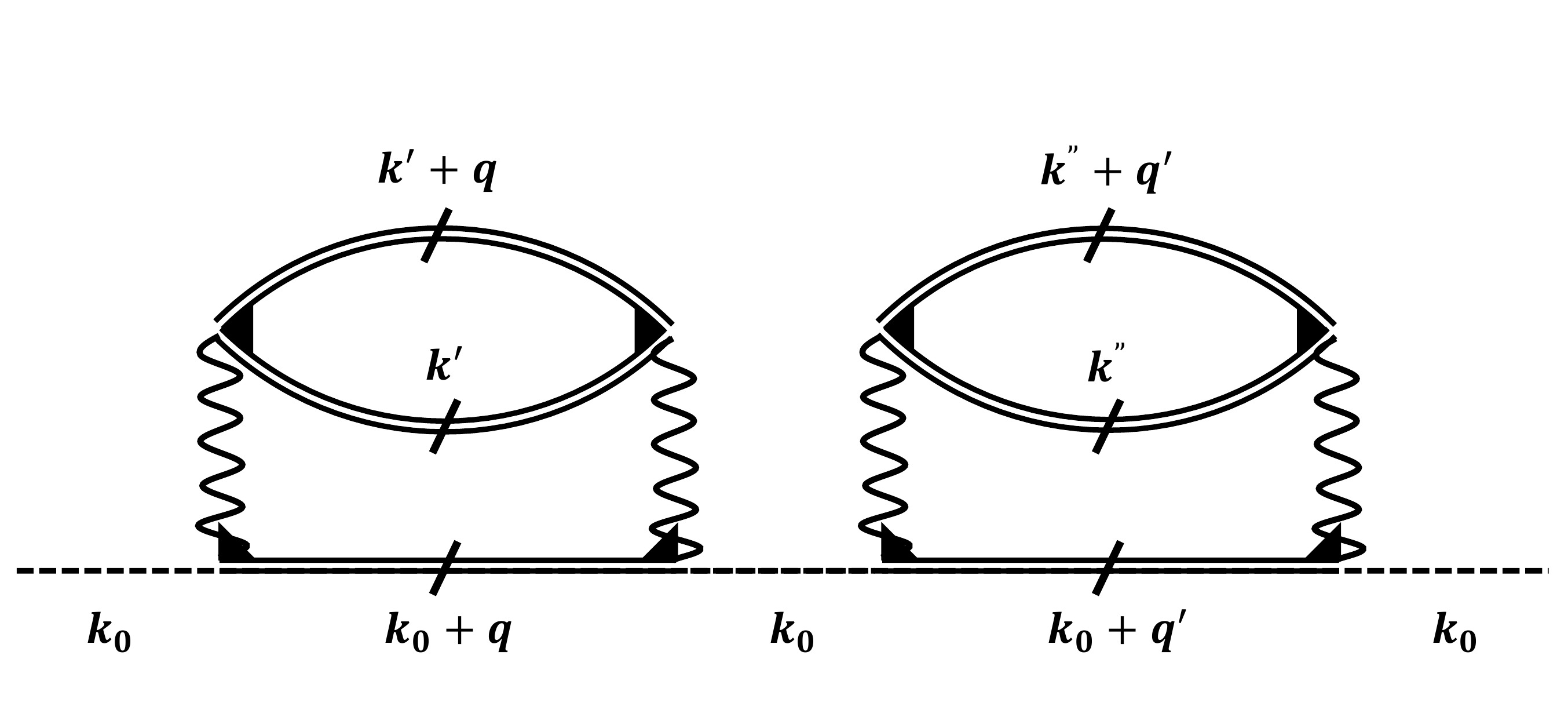}
\caption{Improper forth order Feynman diagram\label{fig:Forth_unpro}}
\end{figure}

Using 
\begin{eqnarray}
S_l(t) &=& \Sigma^T(t)e^{i\xi t}\nonumber\\
S_g(t) &=&( \Sigma^R(t)-\Sigma^T(t)\bigr)e^{i\xi t}\nonumber,
\end{eqnarray}
we can describe the quantum Master equation in second order using the self-energy, which reads
\begin{eqnarray}
\frac{\partial}{\partial t}\rho^I_S(t)&=&\int_{0}^{t}ds \label{retarded}\\ 
&\times \Biggl[& -iS^r_l(t-s)[\mathcal{C}^{\dagger}_{\sigma}\mathcal{C}_{\sigma},\rho^I_S(s)]\nonumber\\
&+&iS^r_g(t-s)[\mathcal{C}_{\sigma}\mathcal{C}^{\dagger}_{\sigma},\rho^I_S(s)]\nonumber\\
&+& S^i_l(t-s)\Bigl(\{\mathcal{C}^{\dagger}_{\sigma}\mathcal{C}_{\sigma},\rho^I_S(s)\}-2\mathcal{C}_{\sigma}\rho^I_S(s)\mathcal{C}^{\dagger}_{\sigma}\Bigr)\nonumber\\
 &+& S^i_g(t-s)\Bigl(\{\mathcal{C}_{\sigma}\mathcal{C}^{\dagger}_{\sigma},\rho^I_S(s)\}-2\mathcal{C}^{\dagger}_{\sigma}\rho^I_S(s)\mathcal{C}_{\sigma}\Bigr)\Biggr]\nonumber\\
S_l(t) &=& S^r_l(t) + iS^i_l(t)\\
S_g(t)& =& S^r_g(t) + iS^i_g(t).
\end{eqnarray}

We also confirm the forth order term of $\mathcal{K}(t)$ which corresponds to the forth order Feynman diagrams.
The forth order term of $\mathcal{K}(t)$ can be written as,
\begin{eqnarray*}
&\mathcal{K}_4(t)=\int_0^t dt_1\int_0^{t_1} dt_2\int_0^{t_2} dt_3 \Bigl(\mathcal{P}\mathcal{L}(t)\mathcal{L}(t_1)\mathcal{L}(t_2)\mathcal{L}(t_3)\mathcal{P}\nonumber\\
& \ \ -\mathcal{P}\mathcal{L}(t)\mathcal{L}(t_1)\mathcal{P}\mathcal{L}(t_2)\mathcal{L}(t_3)\mathcal{P}
\end{eqnarray*}
The first term on the right-hand side corresponds to a forth order Feynman diagram. The other terms can be described by improper Feynman diagrams, such as shown in Fig.~\ref{fig:Forth_unpro}.
The projection operator, $\mathcal{Q}$, in Eq.~(\ref{Pro6}) removes all improper Feynman diagrams from $\mathcal{K}$. 
Higher order terms, $\mathcal{K}_{n}$, can be described in the same way by higher-order terms of the self-energy. Therefore, we can conclude that the self-energy describes the dynamics of the master equation.

In the main text, we use the dynamical mean field theory to calculate the self-energy, which is then used in the master equation.
We calculate $\Sigma^T$ in dynamical mean field theory by calculating
\begin{align}
&\frac{\sum_{\bm{k^{\prime}},\bm{q}}n(\bm{k^{\prime}+q},\omega)(1-n(\bm{k^{\prime}},\omega))(1-n(\bm{k_0+q},\omega))}{\sum_{\bm{k^{\prime}},\bm{q}}(1-n(\bm{k^{\prime}+q},\omega))n(\bm{k^{\prime}},\omega)n(\bm{k_0+q},\omega)}\nonumber\\
& \ \ \ \ \ \ = \Sigma^T(\omega)/(\Sigma^R(\omega)-\Sigma^T(\omega))\label{T}
\end{align}
Eq.~(\ref{T}) is satisfied when the equilibrium state of the bath is described by $\rho_B=\sum_n\ket{n}\bra{n}e^{-\beta\epsilon_n}/\sum_ne^{-\beta\epsilon_n}$.

\section{S3. \ \  Self-energy in density matrix representation}
In this section, we further clarify why $\mathrm{tr_B}\Bigl[\mathcal{B}_{\sigma}(t)\mathcal{B}^{\dagger}_{\sigma}(s)\rho_B\Bigr]$ an similar terms arising in Eq.~(\ref{Pro6}) correspond to the self-energy. The single-particle Green's function $G^T_{\bm{k_0}}(t,s)$ for $t>s$ is defined as
\begin{equation}
    G^T_{\bm{k_0}}(t,s)=\mathrm{tr_{total}}\Bigl[c_{\bm{k_0}}(t)c^{\dagger}_{\bm{k_0}}(s)\rho_{total}\Bigr],
\end{equation}
where $c_{\bm{k_0}}(t)=e^{iHt}c_{\bm{k_0}}e^{-iHt}$. Splitting the Hamiltonian into a free part and the interaction, $H=H_0+H_{int}$, we can use the interaction representation and write
\begin{eqnarray*}
    e^{-iHt}&=&e^{-iH_0t}\mathcal{S}(t)\\
    \mathcal{S}(t)&=&T\exp \bigl[-i\int_0^t ds H_{int}(s)\bigr]\\
    e^{iHt}&=&\mathcal{S}^{-1}(t)e^{iH_0t}\\
    \mathcal{S}(t,s)&=& T\exp\bigl[-i\int_s^t ds_1 H_{int}(s_1)\bigr]=\mathcal{S}(t)\mathcal{S}^{-1}(s)\\
  \mathcal{H}^{\prime}_B&=&  \mathcal{Q}H_{int}\mathcal{Q}\\
  \mathcal{H}_c&=&  \mathcal{P}H_{int}\mathcal{Q}+\mathcal{Q}H_{int}\mathcal{P} \\
    H_{int}&=&\mathcal{H}^{\prime}_B+\mathcal{H}_c\\
    \mathcal{S}_B(t)&\equiv& T\exp\bigl[-i\int_0^t ds \mathcal{H}^{\prime}_B(s)\bigr].
\end{eqnarray*}

By using these relations, we can write
\begin{widetext}
\begin{eqnarray}
    G^T_{\bm{k_0}}(t,s)&=&\mathrm{tr_{total}}\Bigl[\mathcal{S}^{-1}(t)c^{I}_{\bm{k_0}}(t)\mathcal{S}(t-s)c^{I\dagger}_{\bm{k_0}}(s)\mathcal{S}(s)\rho_{tot}\Bigr]\label{total}\\
    &\simeq& \int\int ds_1 ds_2 \mathrm{tr_{total}}\Bigl[\mathcal{S}^{-1}_B(t)c^{I}_{\bm{k_0}}(t)\mathcal{S}_B(t-s_2)\mathcal{H}^I_c(s_2) \times \mathcal{S}_B(s_2-s_1) \mathcal{H}^I_c(s_1)\mathcal{S}_B(s_1-s) c^{I\dagger}_{\bm{k_0}}(s)\mathcal{S}_B(s)\rho_{tot}\Bigr]\label{2nd}\\
    &=& \int\int ds_1 ds_2G^{0T}_{\bm{k_0}}(t,s_2)\mathrm{tr_{B}}\Bigl[\mathcal{B}(s_2)\mathcal{B}^{\dagger}(s_1)\rho_B\Bigr]G^{0T}_{\bm{k_0}}(s_1,s)\\
    &=& \int\int ds_1 ds_2G^{0T}_{\bm{k_0}}(t,s_2)\Sigma^T(s_2-s_1)G^{0T}_{\bm{k_0}}(s_1,s).
\end{eqnarray}
\end{widetext}
When deriving Eq.(\ref{2nd}) from Eq.(\ref{total}), we have used  second order perturbation in $\mathcal{H}_c$. In this equation, $\Sigma^T$ is the 0th order term in $\mathcal{H}_c$, but exact in $\mathcal{H}^{\prime}_B$.
We thus have shown that $\mathrm{tr_B}\Bigl[\mathcal{B}_{\sigma}(t)\mathcal{B}^{\dagger}_{\sigma}(s)\rho_B\Bigr]$ in Eq.(9) in the main text corresponds to the self-energy.

\section{ S4. \ \  Spectral function in the case of larger system}
When considering a system which includes the Hilbert space spanned by $c^{(\dagger)}_{\bm{k_0},\uparrow}$ and $c^{(\dagger)}_{\bm{k_0},\downarrow}$,  additional terms such as $\frac{U}{N}\sum_{\bm{q}}\Bigl(c^{\dagger}_{\bm{k_0}\sigma}c_{\bm{k_0}+\bm{q}\sigma}c^{\dagger}_{\bm{k}+\bm{q}\bar{\sigma}}c_{\bm{k_0}\bar{\sigma}}+ h.c.\Bigr)$ and $\frac{U}{N}\sum_{\bm{q}}\Bigl(c^{\dagger}_{\bm{k_0}\sigma}c_{\bm{k_0}+\bm{q}\sigma}c^{\dagger}_{\bm{k_0}\bar{\sigma}}c_{\bm{k_0}-\bm{q}\bar{\sigma}}+ h.c.\Bigr)$ appear in the coupling Hamiltonian $\mathcal{H}_C$. This leads to additional gain and loss modes, which can be described by the two-particle self-energy, which is however ignored in this paper for simplicity. We can write down the quantum master equation for the density operator of the spectral function, 
\begin{displaymath}
\rho^{SF}(t) = a(t)\ket{\uparrow}\bra{0} + b(t) \ket{2}\bra{\downarrow}, 
\end{displaymath}
where $\ket{0}$ is the unoccupied system and  $\ket{2}=c^{\dagger}_{\bm{k_0},\uparrow}c^{\dagger}_{\bm{k_0},\downarrow}\ket{0}$.
The quantum master equation becomes
\begin{eqnarray}
 &&  \frac{\partial}{\partial t}
    \begin{pmatrix}
      a(t) \\
      b(t) 
    \end{pmatrix}
= \int_{t_0}^{t}ds \times\nonumber\\
&&      \begin{pmatrix}
      S_{eff}(t\mathalpha{-}s)\mathalpha{+}2S^i_g(t\mathalpha{-}s) & \mathalpha{-}2S^i_l(t\mathalpha{-}s) \\
      \mathalpha{-}2S^i_g(t\mathalpha{-}s) & S_{eff}(t\mathalpha{-}s)\mathalpha{+}2S^i_l(t\mathalpha{-}s) 
    \end{pmatrix}
      \begin{pmatrix}
      a(s) \\
      b(s) 
    \end{pmatrix}
  ,\nonumber
  \end{eqnarray}
  Thus, the spectral function is given as
  \begin{eqnarray}
  \frac{\partial}{\partial t}\mathrm{Tr}\Bigl[\mathcal{C}_{\uparrow}\rho^{SF}(t)\Bigr] &=&\frac{\partial}{\partial t}\bigl(a(t) + b(t)\bigr)\\
  &=& \int_{t_0}^{t}ds S_{eff}(t-s)\bigl(a(s)+b(s)\bigr).\nonumber
\end{eqnarray}

In this case, the dynamics of $\rho^{SF}$ originally includes gain and loss modes. However, when calculating the trace for the spectral function, the gain and loss terms disappear. We finally can derive the spectral function by Fourier transformation
\begin{equation}
  A_{\uparrow\uparrow}(\omega)= \frac{1}{\pi}\mathrm{Tr}\bigl[\mathcal{C}\rho^{SF}_S(\omega)\bigr] = i/\pi(\omega-\xi-\Sigma^R(\omega)),
\end{equation}
where we use $a(0)+b(0)=1$ as initial condition. Therefore, even when we consider a larger system, the gain and loss modes appearing in the dynamics of $\rho^{SF}$ cancel in the spectral function, whose dynamics is described by an effective non-hermitian Hamiltonian, $\mathcal{H}_{eff}(\omega)=\mathcal{H}_{0}+\Sigma^R(\omega)$. This statement holds generally in  OQS.

\section{S5. \ \  \label{SFinLindblad}Spectral function in the steady state of open quantum systems}
Here, we verify that gain and loss terms in the quantum Master equation do not affect the dynamics of the spectral function. Therefore, the spectral function can be described by an effective non-hermitian Hamiltonian, which is identical to the effective non-hermitian Hamiltonian in the quantum Master equation under postselection.

We suppose that the density matrix in the steady state is given as $\rho_{SS}=\sum_n a_n\ket{n}\bra{n}$ and that odd powers of the coupling Hamiltonian 
vanish when tracing out the bath, which can be written as $\mathrm{Tr_B}[\mathcal{B}^{2m+1}\rho_B]=0$.  $\mathcal{B}$ is a fermionic operator of the coupling Hamiltonian on the Hilbert space of the bath.
Furthermore, we suppose the absence of gain and loss modes such as $\psi_{\alpha}\rho\psi^{\dagger}_{\beta}$($\alpha\neq\beta$), where $\alpha$ and $\beta$ correspond to an internal quantum numbers such as spin or orbital, and  $\psi_{\alpha}$ is the fermionic annihilation operator of an electron in state $\alpha$. 

In general, the density matrix of a fermionic system can be written as
\begin{eqnarray}
    \rho(t)& =& \sum_{s_{\alpha},s^{\prime}_{\alpha},s_{\beta},s^{\prime}_{\beta},\dots=0,1}D_{s_{\alpha},s^{\prime}_{\alpha},s_{\beta},s^{\prime}_{\beta},\dots}(t)\nonumber\\
    &&\times \psi_{\alpha}^{\dagger (s_\alpha)}\psi_{\beta}^{\dagger (s_\beta)}\dots\ket{0}\bra{0}\dots\psi_{\beta}^{(s^{\prime}_\beta)}\psi_{\alpha}^{(s^{\prime}_\alpha)}\\
     & =& \sum_{n,n^{\prime}}D_{n,n^{\prime}}(t) \ket{n}\bra{n^{\prime}},
\end{eqnarray}
where $n^{(\prime)}=(s_{\alpha}^{(\prime)},s_{\beta}^{(\prime)},\dots)$ and $s_{\alpha}^{(\prime)}$ represents the occupation number of a particle in state $\alpha$. 

We consider the spectral function $A_{\alpha\alpha}(t) = \mathrm{Tr}[(\psi_{\alpha}(t)\psi^{\dagger}_{\alpha}+\psi^{\dagger}_{\alpha}\psi_{\alpha}(t))\rho_{SS}]$, which can be written using
$\rho^{SF}=(\psi^{\dagger}_{\alpha}\rho_{SS}+\rho_{SS}\psi^{\dagger}_{\alpha})$.
The density matrix of the spectral function and its initial condition can be written using $m=(s_{\beta},s_{\gamma},\dots)$
\begin{align}
    &\rho^{SF}(t) = \sum_{m} B_{m}(t)\psi_{\alpha}^{\dagger}\ket{m}\bra{m}\\
    &\sum_m B_{m}(0) = 1\\
 &\Rightarrow   A_{\alpha\alpha}(t)=\mathrm{Tr}\Bigl[\psi_{\alpha}\rho^{SF}(t)\Bigr] = \sum_{m} B_{m}(t).
\end{align}
We consider now the contribution to $\frac{\partial}{\partial t}\sum_m B_m(t)$ from  the gain and loss terms $S^l_{\beta}(t-s)\psi_{\beta}\rho^{SF}(s)\psi_{\beta}^{\dagger}$ ($\beta\neq\alpha$). It is important to see that
the  (non-hermitian) counterpart, $-\frac{1}{2}S^l_{\beta}(t-s)\{\psi_{\beta}^{\dagger}\psi_{\beta},\rho^{SF}(s)\}$, must exist for each gain and loss term, 
because of the  conservation of probability in the dynamics of the quantum master equation.
By defining $\psi^{\dagger}_{\alpha}\ket{m_1}\bra{m_1}=\psi^{\dagger}_{\alpha}\psi^{\dagger}_{\beta}\ket{m_2}\bra{m_2}\psi_{\beta}$, we can rewrite this part of the master equation for the spectral function as
\begin{eqnarray}
    &&S^l_{\beta}(t-s)\Bigl[\psi_{\beta}\psi^{\dagger}_{\alpha}\ket{m_1}\bra{m_1}\psi^{\dagger}_{\beta} - \frac{\bigl\{\psi^{\dagger}_{\beta}\psi_{\beta},\psi^{\dagger}_{\alpha}\ket{m_1}\bra{m_1}\bigr\}}{2}\Bigr]\nonumber\\
     &=& S^l_{\beta}(t-s) \Bigl[\psi^{\dagger}_{\alpha}\ket{m_2}\bra{m_2}-\psi^{\dagger}_{\alpha}\ket{m_1}\bra{m_1}\Bigr].\label{GL}
\end{eqnarray}
If we take the trace,  $\mathrm{Tr[\psi_{\alpha}\sim]}$, the contribution of the loss term and the counterpart vanish. The arguments above hold true for any ($\beta,m$) and for the gain terms. We note that to derive Eq.~(\ref{GL}), we have to consider the commutation relation between $\mathcal{H}_c$ and $\psi^{\dagger}_{\alpha}$.

On the other hand, gain and loss terms including $S^{g(l)}_{\alpha}$ vanish because $\psi_{\alpha}\rho^{SF}\psi^{\dagger}_{\alpha} = \psi^{\dagger}_{\alpha}\rho^{SF}\psi_{\alpha} =0$ as is written in the main text. 
The spectral function $A_{\alpha\alpha}$ is not affected by $S^{l/g}_\beta$ and can be described only by the non-hermitian term $S^{R}_{\alpha}$. Therefore, the spectral function of a general fermionic OQS can be described by an effective non-hermitian Hamiltonian without postselection.

Although we here have assumed that  gain and loss modes such as $\psi_{\alpha}\rho\psi^{\dagger}_{\beta}$($\alpha\neq\beta$) do not exist, we will show in the next section that the spectral function in the periodic Anderson model(PAM) is also only described by a non-hermitian Hamiltonian. In the PAM, such gain and loss terms appear due to the hybridization between the conduction- and the $f$-electrons.

\section{S6. \ \  Quantum master equation in the periodic Anderson model} 
In the main text, we have proven in the Hubbard model that the non-hermitian  Hamiltonian describing the spectral functions is identical to the non-hermitian Hamiltonian in the quantum master equation under postselection. Here, we extend our considerations to the periodic Anderson model(PAM) reading
\begin{eqnarray}
\mathcal{H}_{\mathrm{PAM}}&=&\sum_{\bm{k}\sigma}\Bigl((\epsilon_{\bm{k}}+\mu_c)c^{\dagger}_{\bm{k}\sigma}c_{\bm{k}\sigma}
 \mathalpha{+} (\epsilon_{f\bm{k}}+\mu_f) f^{\dagger}_{\bm{k}\sigma}f_{\bm{k}\sigma}\nonumber\\
&& \mathalpha{+}V(f^{\dagger}_{\bm{k}\sigma}c_{\bm{k}\sigma}\mathalpha{+}h.c.)\Bigr)\mathalpha{+}U \sum_i n^f_{i\uparrow}n^f_{i\downarrow}\label{X}
\end{eqnarray}
where $c^{(\dagger)}_{\bm{k}\sigma},f^{(\dagger)}_{\bm{k}\sigma}$ are annihilation (creation) operators of the $c$- and the $f$-electrons for momentum $\bm{k}$ and spin-direction $\sigma$. $\epsilon_{c,f}$ are the kinetic energy for the $c$- and the $f$-electrons, $\mu_{c,f}$ the chemical potentials for the $c$- and $f$-electron band, $V$ a local hybridization, and $U$ a density-density interaction for the $f$ electrons.

As in the Hubbard model, we first divide the total Hamiltonian into the system, the bath, and the coupling Hamiltonian,
\begin{eqnarray}
H&=&H_S+H_B+H_c\\
\mathcal{H}_S&=&(\epsilon_{\bm{k_0}}\mathalpha{+}\mu_c)c^{\dagger}_{\bm{k_0}\sigma}c_{\bm{k_0}\sigma}
 \mathalpha{+}(\epsilon_{f\bm{k_0}}\mathalpha{+}\mu_f\mathalpha{+}Un_{\bar{\sigma}}) f^{\dagger}_{\bm{k_0}\sigma}f_{\bm{k_0}\sigma}\nonumber\\
&& \mathalpha{+}V(f^{\dagger}_{\bm{k}\sigma}c_{\bm{k}\sigma}\mathalpha{+}h.c)\\
 &\mathalpha{=}&\sum_{\pm}\epsilon_{\pm}\xi^{\dagger}_{\pm,\sigma}\xi_{\pm,\sigma}\\
\mathcal{H}_B&=&\sum_{\bm{k}\neq\bm{k_0}}\Bigl((\epsilon_{\bm{k}}+\mu_c)c^{\dagger}_{\bm{k}\sigma}c_{\bm{k}\sigma}
 \mathalpha{+} (\epsilon_{f\bm{k}}+\mu_f) f^{\dagger}_{\bm{k}\sigma}f_{\bm{k}\sigma}\\
&& \mathalpha{+}V(f^{\dagger}_{\bm{k}\sigma}c_{\bm{k}\sigma}\mathalpha{+}h.c)\Bigr)\nonumber\\
&&  \mathalpha{+}U\sum_{\bm{k_1},\bm{k_2},\bm{k_3},\bm{k_4}\neq\bm{k_0}}\delta_{\bm{k_1}+\bm{k_3},\bm{k_2}+\bm{k_4}}f^{\dagger}_{\bm{k_1}\sigma}f_{\bm{k_2}\sigma}f^{\dagger}_{\bm{k_3}\bar{\sigma}}f_{\bm{k_4}\bar{\sigma}}\nonumber\\
\mathcal{H}_c&=&U\sum_{\bm{k_0},\bm{k_1},\bm{k_2},\bm{k_3}}\delta_{\bm{k_1}+\bm{k_3},\bm{k_0}+\bm{k_2}}\Bigl(f^{\dagger}_{\bm{k_0}\sigma}f_{\bm{k_1}\sigma}f^{\dagger}_{\bm{k_2}\bar{\sigma}}f_{\bm{k_3}\bar{\sigma}}+ h.c.\Bigr),\nonumber\\
\end{eqnarray}
The Hilbert space of the system contains the $c$-electron  in $(\bm{k_0},\sigma)$ and the $f$-electron in $(\bm{k_0},\sigma)$, where the spin-direction $\sigma$ is fixed.

 Although the size of the Hilbert space of the system becomes larger than in the Hubbard model, we can derive the QME in the same way, ignoring the two-particle self-energy.
 The quantum master equation is given as 
 \begin{eqnarray}
&&\frac{\partial}{\partial t}\rho^I_S(t)=\int_{t_0}^{t}ds\times\nonumber\\ 
&&\Bigl\{-i\Bigl[(\xi_{-\sigma}^{\dagger} \ \xi_{+\sigma}^{\dagger})
\begin{pmatrix}
      S^R_{r--}(t-s) & S^R_{r-+}(t-s) \\
      S^R_{r+-}(t-s) & S^R_{r++}(t-s) \\
    \end{pmatrix}
      \begin{pmatrix}
      \xi_{-\sigma} \\
      \xi_{+\sigma} \\
    \end{pmatrix}
    , \rho^I_S(s)\Bigr]\nonumber\\
&& +S^l_{i\alpha\beta}(t-s)\Bigl(\bigl\{\xi_{\alpha}^{\dagger}\xi_{\beta},\rho_S(s)\bigr\}-2\xi_{\beta}\rho_S(s)\xi_{\alpha}^{\dagger}\Bigr)\nonumber\\
&& +S^g_{i\alpha\beta}(t-s)\Bigl(\bigl\{\xi_{\alpha}\xi_{\beta}^{\dagger},\rho_S(s)\bigr\}-2\xi^{\dagger}_{\beta}\rho_S(s)\xi_{\alpha}\Bigr)\Bigr\}
 \end{eqnarray}
 with
 \begin{eqnarray*}
S^R_{r\alpha\beta}(t-s) &=& s_{\alpha}s_{\beta}\mathrm{Re}(\Sigma^R(t)e^{i(\xi_{\alpha}t-\xi_{\beta}s)})\\
S^{l}_{\alpha\beta}(t-s)&=&s_{\alpha}s_{\beta}\mathrm{Im}\bigl(\Sigma^T(t)e^{i(\xi_{\alpha}t-\xi_{\beta}s)}\bigr)\\
S^{g}_{\alpha\beta}(t-s)&=&s_{\alpha}s_{\beta}\mathrm{Im}\bigl((\Sigma^R(t)-\Sigma^T(t))e^{i(\xi_{\alpha}t-\xi_{\beta}s)}\bigr)\\
f^{(\dagger)}_{\bm{k_0}\sigma} &=& -s_- \xi_{-\sigma}^{(\dagger)} + s_{+} \xi_{+\sigma}^{(\dagger)}, \ c^{(\dagger)}_{\bm{k_0}\sigma} = s_+ \xi_{-\sigma}^{(\dagger)} + s_- \xi_{+\sigma}^{(\dagger)}\\ 
s_{\pm}&=& \Bigl\{\bigl(\sqrt{h^2_1+V^2}\pm h_1 \bigr)/ 2\sqrt{h^2_1+V^2}\Bigr\}^{\frac{1}{2}}\\
h_1&=&\bigl(\epsilon_{\bm{k}}-\epsilon_{f\bm{k}}\bigr)/2
\end{eqnarray*}

As in the Hubbard model, when fixing the particle number and the magnetization of the system by postselection, we find that the dynamics of the density matrix (under postselection)  
in the limit of $t_0\rightarrow -\infty$ and using the  Markov approximation is given by an effective non-hermitian Hamiltonian as
\begin{equation}
\frac{\partial}{\partial t}\rho^{PS}_S(t)=-i\Bigl(\mathcal{H}_{eff}\rho^{PS}_S(t)-\rho^{PS}_S(t)\mathcal{H}^{\dagger}_{eff}\Bigr)\\
\end{equation}
where the effective non-hermitian Hamiltonian is given as
\begin{eqnarray}
\mathcal{H}_{eff}&=&\epsilon_c(\bm{k_0})c_{\sigma}^{\dagger}c_{\sigma}+V(c_{\sigma}^{\dagger} f_{\sigma} + h.c.)\nonumber\\
&& +\Bigl(\epsilon_f(\bm{k_0}) + \Sigma^R \Bigr) f_{\sigma}^{\dagger}f_{\sigma}\label{PAM_Heff}\\
\Sigma^R&=&\frac{\Sigma^R(\xi_+)+\Sigma^R(\xi_-)}{2} \nonumber\\&&+\frac{h_1(\Sigma^R(\xi_+)-\Sigma^R(\xi_-))}{2\sqrt{h_1^2+V^2}}.\label{SE_PAM}
\end{eqnarray}
We here have used postselection as ($\xi_{\pm\uparrow}^{\dagger}\xi_{\pm\uparrow}=\xi_{\pm\downarrow}^{\dagger}\xi_{\pm\downarrow}=0.5 \leftrightarrow \xi_{\pm\uparrow}^{\dagger}\xi_{\pm\uparrow}=\xi_{\pm\uparrow}\xi_{\pm\uparrow}^{\dagger}$) to derive Eq.~(\ref{PAM_Heff}). We suppose that the observation under postselection leads to  $\xi_{\pm\uparrow}^{\dagger}\xi_{\pm\uparrow}+\xi_{\pm\downarrow}^{\dagger}\xi_{\pm\downarrow}=1$ and that there is no magnetization. 

Here, we have derived the effective non-hermitian Hamiltonian in the context of OQS. 
The dynamics is again described by the retarded self-energy. If we ignore the frequency dependence of the self-energy ($\Sigma^R(\xi_+)=\Sigma^R(\xi_-)=\Sigma^R$), we see that the 
effective non-hermitian Hamiltonian describing the dynamics in the quantum master equation and the effective non-hermitian  Hamiltonian of the spectral function agree with each other. 

In the case of non-Markovian dynamics, the time-dependence of the self-energy must be considered, which makes an analytical comparison between the density matrix under postselection and the spectral function difficult.
Therefore, we numerically compare the non-Markovian dynamics of the quantum master equation with those of the single-particle Green's function calculated by the DMFT/NRG.

\section{S7. \ \  Dynamics of the PAM in the quantum Master equation}
\begin{figure}[t]
\includegraphics[width=0.98\linewidth]{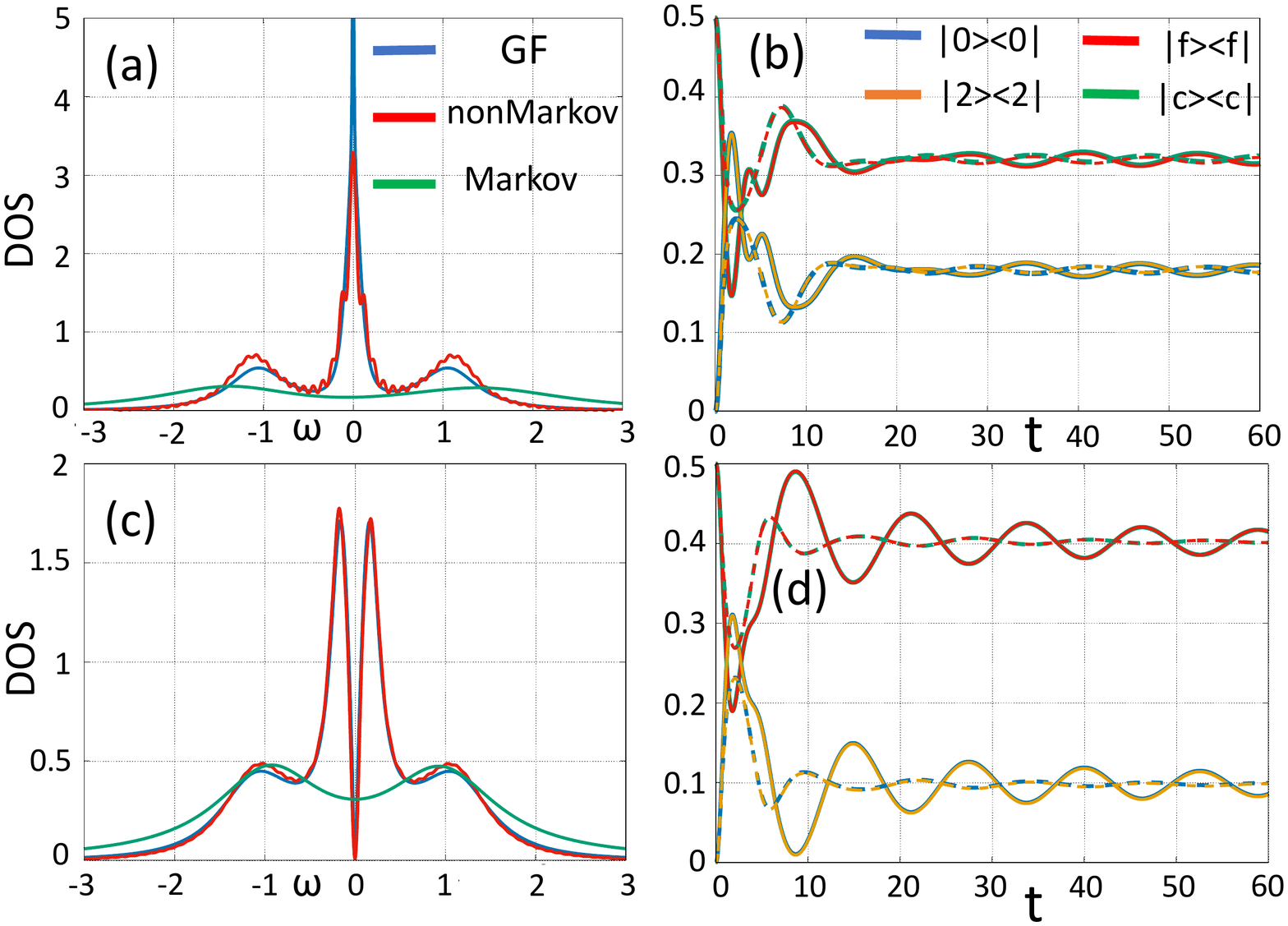}
\caption{Spectral functions and matrix elements of the density matrix in the metallic regime at high temperature and the Kondo insulating phase at low temperature.
The parameters in (a) and (b) are as follows:  $t_c=1.0, t_f=-0.05$, $\mu_c=0, \mu_f=-1.0$, $U=2.0$, $V=0.36,$ and the temperature is $T=0.13$. The parameters in (c) and (d) are $t_c=1.0, t_f=-0.05$, $\mu_c=0, \mu_f=-1.0$, $U=2.0$, $V=0.5$ and $T=0.0002$. The blue, red and green lines in (a) and (c) show respectively the spectral functions calculated directly by the Green's function, the spectral function using non-Markovian dynamics and the Markov dynamics in the limit of $t_0\rightarrow -\infty$. The dashed (full) lines in (b) and (d) show the Markovian dynamics (non-Markovian dynamics) of the diagonal elements of the density matrix from the initial state $\rho_i=\xi^{\dagger}_-|0><0|\xi_-$.\label{fig:PAMSF1}}
\end{figure}

We here compare the Markovian dynamics and the non-Markovian dynamics in the PAM by numerical simulations in the metallic phase above the Kondo temperature and in the Kondo insulating phase below the Kondo temperature. 
We here use the self-energy as calculated by dynamical mean-field theory (DMFT) combined with the numerical renormalization group (NRG) and calculate the spectral function and the diagonal elements of the density matrix using the quantum master equation with and without Markov-approximation.
Figure ~\ref{fig:PAMSF1}(a) and (c) show the spectral functions calculated by the DMFT/NRG and the spectral function calculated by the QME, where $\bm{k_0}=(0.5\pi,0.5\pi)$. 
 
The spectral function calculated by the QME with non-Markovian dynamics agrees with the spectral function calculated directly from the Green's function. This shows that the real time dynamics of the spectral function corresponds to the dynamics of the QME under postselection because gain and loss terms vanish in the spectral function.
Therefore,  the effective non-hermitian Hamiltonian in the spectral function corresponds to the effective non-hermitian Hamiltonian in the non-Markov QME under postselection.
In Fig.~\ref{fig:PAMSF1}(a) and (c), the spectral function by the QME using the Markov approximation in the limit of $t_0\rightarrow -\infty$ only includes the self-energy at $\omega=\xi_\pm$ and  neglects the frequency dependence of the self-energy around the Fermi energy. In this case, the effective Hamiltonian in the QME describes  the scattering away from the Fermi energy. Thus, the spectral function calculated by the QME with Markov approximation includes the particle-hole excitations at $\omega=\pm \frac U2$ but cannot describe the excitation near the Fermi energy.
The peaks at high temperature in Fig.~\ref{fig:PAMSF1}(a) are smeared out wider than those at low temperature due to the stronger scattering at high temperature.

Figures \ref{fig:PAMSF1}(b) and (d), show that the relaxation of the diagonal elements form the initial state $\rho_i=\xi^{\dagger}_-|0><0|\xi_-$ using  Markovian and non-Markovian dynamics. In the metallic regime above the Kondo temperature, Fig.~\ref{fig:PAMSF1}(b), these elements oscillate but are strongly damped.  On the other hand, in the Kondo-insulating regime below the Kondo temperature, the oscillation persists for a long time. We believe that this change of the dynamics is related to the Kondo crossover. 
We note that the Markov dynamics in the limit of $t_0\rightarrow -\infty$ in Fig.~\ref{fig:PAMSF1} does not significantly change between high-temperature metallic state and low-temperature insulating state. Therefore, we can conclude that the Kondo crossover from the metallic behavior at high temperature to the insulator at low temperature originates from the change in the non-Markovian dynamics.

\end{document}